\documentclass[aps,
prb,
preprintnumbers,
superscriptaddress,
footinbib,
amsfonts,
amssymb,
amsmath,
intlimits,
onecolumn,
]{revtex4-2}
\usepackage[utf8]{inputenc}
\usepackage[T2A]{fontenc}
\usepackage{bm,latexsym,mathrsfs,enumerate,amsmath,amssymb}
\usepackage{upgreek}
\usepackage[table,x11names]{xcolor}
\usepackage[breaklinks=true,
unicode=true,
urlcolor = blue,
colorlinks = true,
citecolor = blue,
linkcolor = blue
]{hyperref}
\usepackage{graphicx}
\usepackage{wasysym}
\usepackage[cal=boondoxo,scr=rsfs]{mathalfa}
\usepackage{mathtools}

\usepackage{makecell,tabularx,multirow}
\usepackage{hhline}

\usepackage{chemformula}
\newcommand{\bise}{Bi$_2$Se$_3$}

\usepackage{titlesec}

\titleformat{\section}[display]%
{\raggedright\large\bfseries}
{\thesection}
{1em}
{}
\titlespacing*{\section}
{0pt}
{1.3cm}
{0pt}

\usepackage{easyReview}
\begin{document}

\title{
Giant geometric nonlinear transport in inhomogeneously strained \ch{Bi2Se3} single crystalline slabs
}

\author{Igor~Veremchuk}
\email{i.veremchuk@hzdr.de}
\affiliation{Helmholtz-Zentrum Dresden-Rossendorf e.V., Institute of Ion Beam Physics and Materials Research, 01328 Dresden, Germany}

\author{Pavlo~Makushko}
\affiliation{Helmholtz-Zentrum Dresden-Rossendorf e.V., Institute of Ion Beam Physics and Materials Research, 01328 Dresden, Germany}

\author{Anita~Guarino}
\affiliation{CNR-SPIN, c/o Università di Salerno, via Giovanni Paolo II 132 – Fisciano (Sa), Italy}

\author{Muhammad~Waqee~Ur~Rehman}
\affiliation{CNR-SPIN, c/o Università di Salerno, via Giovanni Paolo II 132 – Fisciano (Sa), Italy}

\author{Rosalba~Tatiana~Fittipaldi}
\affiliation{CNR-SPIN, c/o Università di Salerno, via Giovanni Paolo II 132 – Fisciano (Sa), Italy}

\author{Tobias~Kosub}
\affiliation{Helmholtz-Zentrum Dresden-Rossendorf e.V., Institute of Ion Beam Physics and Materials Research, 01328 Dresden, Germany}

\author{Alexej~Pashkin}
\affiliation{Helmholtz-Zentrum Dresden-Rossendorf e.V., Institute of Ion Beam Physics and Materials Research, 01328 Dresden, Germany}

\author{Fabian~Ganss}
\affiliation{Helmholtz-Zentrum Dresden-Rossendorf e.V., Institute of Ion Beam Physics and Materials Research, 01328 Dresden, Germany}

\author{Oleg~A.~Petrenko}
\affiliation{Department of Physics, University of Warwick, Coventry CV4 7AL, United Kingdom}

\author{Proloy~T.~Das}
\affiliation{Helmholtz-Zentrum Dresden-Rossendorf e.V., Institute of Ion Beam Physics and Materials Research, 01328 Dresden, Germany}

\author{Rui~Xu}
\affiliation{Helmholtz-Zentrum Dresden-Rossendorf e.V., Institute of Ion Beam Physics and Materials Research, 01328 Dresden, Germany}

\author{Rhonald~Burgos~Atencia}
\altaffiliation{Present address: Direcci\'on de Ciencias B\'asicas, Universidad Tecnol\'ogica de Bol\'ivar, Cartagena, Colombia}
\affiliation{Dipartimento di Fisica ``E. R. Caianiello", Universit\`a di Salerno, IT-84084 Fisciano (SA), Italy}

\author{Debottam~Mandal}
\affiliation{Dipartimento di Fisica ``E. R. Caianiello", Universit\`a di Salerno, IT-84084 Fisciano (SA), Italy}

\author{Antonio~Vecchione}
\email{antonio.vecchione@spin.cnr.it}
\affiliation{CNR-SPIN, c/o Università di Salerno, via Giovanni Paolo II 132 – Fisciano (Sa), Italy}

\author{Carmine~Ortix}
\email{cortix@unisa.it}
\affiliation{CNR-SPIN, c/o Università di Salerno, via Giovanni Paolo II 132 – Fisciano (Sa), Italy}
\affiliation{Dipartimento di Fisica ``E. R. Caianiello", Universit\`a di Salerno, IT-84084 Fisciano (SA), Italy}

\author{Denys~Makarov}
\email{d.makarov@hzdr.de}
\affiliation{Helmholtz-Zentrum Dresden-Rossendorf e.V., Institute of Ion Beam Physics and Materials Research, 01328 Dresden, Germany}

\begin{abstract}
Electrical transport in non-centrosymmetric materials exhibits a nonlinear conductivity tensor component that generates second-harmonic voltages in response to applied electric fields. Harnessing this effect is essential for developing zero-threshold electromagnetic energy harvesters and broadband frequency upconverters, both of which require maximizing the nonlinear responsivity. Here, we demonstrate that inhomogeneously strained macroscopic slabs of the topological insulator \ch{Bi2Se3} achieve a record-high nonlinear responsivity of $10^4\,$µm/V: a six orders of magnitude boost
compared to its thin-film counterparts. We attribute this giant enhancement to a classical geometric nonlinear transport effect, which is driven by complex strain gradients within the bulk material. Notably, a similar geometric mechanism known as geometric nonlinear Hall was previously observed only in microscale-patterned structures. Our findings establish macroscopic strain engineering as a powerful route to enhance nonlinear transport, with immediate relevance to a broad class of layered systems, including transition metal dichalcogenides and moir\'e heterostructures.
\end{abstract}

\maketitle	

\section*{Introduction}

Second-order electrical transport responses have attracted intensive attention due to their strong technological relevance for frequency multipliers and energy harvesters. Pioneering works on the nonlinear Hall effect in non-centrosymmetric quantum materials have predicted remarkably high second-order conversion efficiencies of up to 13\% at zero bias and zero threshold, which is a major advantage over contemporary semiconductor-based upconverters and rectifiers.

Inspired by these theoretical predictions, numerous studies have investigated nonlinear transport in 
low symmetric
materials, where the responses are enabled by 
a
band geometric property of Bloch electrons known as
Berry curvature dipole (BCD). However, due to strict symmetry requirements, the portfolio of suitable materials remains limited, and their operation is typically restricted to cryogenic temperatures [Table~\ref{tab:responsivity}]. To overcome these limitations, two primary approaches have been developed to enable second-order electric processes in the largest class of known materials, namely nonmagnetic systems with spatial inversion symmetry. On the one hand, second-order transport in these systems can arise from 
surface states which inherently break inversion symmetry. The corresponding nonlinear conductivity can be either activated by a surface BCD or, at surfaces with threefold rotation symmetry, by a higher moment of the Berry curvature, dubbed the Berry-curvature triple (BCT)~\cite{Du_2019,Isobe_2020,Ortix_2021,Tiwari_2021,He_2022,Duan_2022}. This mechanism has been observed in the three-dimensional topological insulators Bi$_2$Se$_3$~\cite{He_2021} and Bi$_2$Te$_3$~\cite{Shen_2025} and in elemental bismuth at room temperature~\cite{Makushko_2024}.
Complementary to these 
mechanisms related to the geometric properties of the electronic wavefunctions,
second-order responses 
can be also observed in
any classical conducting system with structurally broken inversion symmetry, as exemplified by arc-shaped graphene~\cite{Schade_2019} and Bi microstructures~\cite{Makushko_2024}. By analogy to, and inspired by its quantum counterpart, this phenomenon in curvilinear conductors is known as the geometric nonlinear Hall effect (gNLHE).

To date, significant progress has been made in 
optimizing
the performance of nonlinear Hall materials, with the highest reported nonlinear responsivities already exceeding $>1000\,$µm/V  [Table~\ref{tab:responsivity}].
Here, we report on a giant nonlinear responsivity in inhomogeneously strained \bise{} single crystals, where this application-relevant performance metric exceeds $10^4\,$µm/V for the first time. This record value is higher than that of any previously reported nonlinear Hall material and represents a six-order-of-magnitude enhancement compared to thin films of \ch{Bi2Se3}. The underlying inhomogeneous strain distribution is experimentally realized by the multiple cleavage of macroscopic single crystalline slabs of \ch{Bi2Se3}. 
The giant enhancement with respect to TI thin films, where skew scattering parametrized by the BCT~\cite{He_2021} governs nonlinear transport, 
implies the occurrence of a different mechanism unrelated to
the quantum geometry of the Bloch waves.
We argue that this effect is purely classical and stems
from macroscopic inversion symmetry breaking within the bulk of the crystal, induced by the strain gradient. 
This effect thus represents a strain-driven counterpart of the gNLHE yet it enables nonlinear transport in any in-plane direction. Therefore, we refer to this effect as the geometric nonlinear transport effect (gNLTE).
We demonstrate that the gNLTE yields highly efficient
second-harmonic processes highlighting 
its potential
in frequency upconversion and rectification.

\begin{table}[]
\caption{\textbf{Comparison of nonlinear responsivities for different material families.}
The table summarizes the maximal values of nonlinear responsivities measured at the indicated temperature. Furthermore, we indicate maximal temperatures, $T_\text{max}$, at which the nonlinear response is reported. The origin of the nonlinear effect and the type of the sample are stated as well.
Responsivity values indicated with $^*$ ($^{**}$) were measured under applied magnetic field (electric gate bias voltage).
}
\label{tab:responsivity}
\begin{tabular}{|l|l|l|l|l|l|}
\hline
Material                & $T_\text{max}$ (K)  & $E_{yxx}/(E_{xx}^2)$ (µm/V)                                   & $E_{xxx}/(E_{xx}^2)$ (µm/V)                    & Origin                    & Sample type               \\ \hhline{======}
\bise{} slab (this work)                  & 200             & $9.5 \times 10^{3}$ @ 10 K                                                   & $14.6 \times 10^{3}$ @ 10 K                                    & geometric        & Single crystal slab          \\ \hline
graphene arc~\cite{Schade_2019}       & 300                 & $5.6\times10^{-2}$ @300 K          &                             & geometric              &   curved Hallbar     \\ \hline
Bi~\cite{Makushko_2024}                 & 300             & $7.9 \times 10^{-2}$  @ 300 K               &                                     & BCT + geometric   & Thin film, Curved Hallbar \\ \hhline{======}
\bise{}~\cite{He_2021}  & 200 & $0.75\times10^{-3}$ @ 20\,K  & & BCT   & Epitaxial film\\ \hline
Bi~\cite{Makushko_2024}                 & 300             & $1.09 \times 10^{-2}$ @ 300 K              &                                     & BCT (disorder mediated)   & Thin film                 \\ \hline
hBN/graphene/hBN~\cite{He_2022}        & 120             & $5.6 \times 10^{2}$ ** @ 1.7 K                 & $2.2 \times 10^{2}$ **  @ 1.7K & Disorder mediated       & Thin   film               \\ \hline
\ch{WTe2}~\cite{Kang2019}                    & 100             & $1.4 \times 10^{-3}$ @ 2 K                       &                                     & BCD + disorder mediated & Flakes                    \\ \hline
t-\ch{WSe2}~\cite{Huang2022}                  & 40              & $3.6$ @ 1.5 K                                          &                                     & BCD                       & twisted   2D layers       \\ \hline
strained WSe2~\cite{Qin2021}         & 140             & $1.4 \times 10^{-1}$ @ 80 K                    &                                     & BCD                       & Strained film             \\ \hline
\ch{Te}~\cite{Suarez_2024}                      & 100             & $8.3\times10^{-1}$ ** @ 10 K                                        & $7.1\times10^{-1}$ ** @ 10K                        & Disorder   mediated       & Thin   film               \\ \hline
\ch{Te}~\cite{Cheng_2024}                      & 300             & $72 \times 10^{-3}$ ** @200 K                       &                                     & Disorder   mediated       & Flakes                    \\ \hline
\ch{Bi2Te3}~\cite{Shen_2025}                  & 300             & $6 \times 10^{-3}$ @ 300 K                         &                                     & BCT   (disorder mediated) & Epitaxial   films         \\ \hline
\ch{MoTe2}~\cite{Ma_2022}                   & 100             & $7 \times 10^{-3}$ @ 5 K                         &                                     & BCD +   disorder mediated & Single   crystal flakes   \\ \hline
\ch{TaIrTe4}~\cite{Kumar_2021}                 & 300             & $2.3 \times 10^{-2}$ @ 2 K                       &                                     & BCD                       & Flakes                    \\ \hline
Td-\ch{MoTe2}~\cite{Tiwari_2021}              & 40              & $2\times10^{-2}$ @ 2 K                       &                                     & Disorder   mediated       & Flakes                    \\ \hline
\ch{BaMnSb2}~\cite{Min_2023}                 & 300             & $15 \times 10^{2}$ @ 200K                         &                                     & BCD                       & Flakes                    \\ \hline
4SL-\ch{MnBi2Te4}~\cite{Wang_2023}            & 20              & $1.04 \times 10^{-2}$ @ 2 K                                   & $1.58$ @ 2 K                          & BNQM                      & Flakes                    \\ \hline
\ch{MnBi2Te4}~\cite{Liu2024}                & 320             & $3 \times 10^{-4}$ @ 40 K                  &                                     & Disorder   mediated       & Film                      \\ \hline
Pt~\cite{Min2024} & 300             & $1.4 \times 10^{3}$ @ 10 K                        &                                     & Disorder   mediated       & FEBID microstructures   \\ \hline
\ch{TbMn6Sn6}~\cite{Zhao_2025}                & 320             & $1.58$ @ 300 K                                        &                                     & BNQM                      & Film                      \\ \hline
\ch{YbCo6Ge6}~\cite{Ain2026} &  300  &   $4.5\times10^{3}$ @80\,K &   &  Disorder mediated & FIB-structured single crystal \\ \hline
\ch{CdGeAs2}~\cite{Lee2025}                 & 350             & $2.2 \times 10^{3}$ @ 300 K                         &                                     & Disorder   mediated       & Single crystal flakes                       \\ \hline
\ch{LaAlO3/SrTiO3}~\cite{Sala_2025}           & 3               &                                                     & $1 \times 10^{2}$ @ 3 K      & BNQM                      & Slab                      \\ \hline
\end{tabular}
\end{table}

\section*{Results and discussion}

\subsection*{Structural properties of \bise{} single-crystalline slabs.}

[$0\,0\,0\,1$]-oriented single-crystalline \bise{} specimens were synthesised using a modified optical floating zone technique~\cite{Guarino2024}. \bise{} crystallizes in the $R\overline{3}m$ space group where covalently bonded quintuple layers (QL) are interconnected by van der Waals (vdW) gaps [Fig.~\ref{fig:fig1}(a)]. Relevant details on the surface morphology, stoichiometry are given in Supplementary Note I, the crystal lattice orientation is determined using electron backscatter diffraction (EBSD) [Supplementary Fig.~1]. The pristine \bise{} crystals [Fig.~\ref{fig:fig1}(b)] were cleaved multiple times to reduce their thickness to $200\,$µm [Fig.~\ref{fig:fig1}(c,d)] and cut into rectangular slabs for electrical transport measurements. Upon cleaving, the crystal typically breaks at the vdW bonds, resulting in thinner bulk topological insulator slabs that preserve the long-range structural coherence of \bise{} [Fig.~\ref{fig:fig1}(d)].  

The structural properties of the \bise{} slab were studied by X-ray diffraction (XRD), including $\theta$-$2\theta$ scans [Fig.~\ref{fig:fig1}(e)], reciprocal space maps (RSM) [Fig.~\ref{fig:fig1}(f), Supplementary Figs.~2,\,4], Lauegrams [Fig.~\ref{fig:fig1}(g), Supplementary Figs.~2,\,3], pole figures [Supplementary Fig.~2,5], and Raman scattering [Fig.~\ref{fig:fig1}(h,i), Supplementary Note I.D].
The XRD data confirm the preserved structural integrity and three-fold rotation symmetry of the consecutively cleaved \bise{} crystal. The RSM of the asymmetric ($0\,1\,\overline{1}\,20$) reflection appears as a narrow streak extending along the circular track [dashed line in Fig.~\ref{fig:fig1}(f)], which results from a rotation of the crystal lattice. Similar features are observed for the RSM of the symmetric ($0\,0\,0\,15$) reflection [Supplementary Fig.~4]. This observation is in line with prior reports on a \bise{} crystal subjected to multiple cleavage steps~\cite{Cavallin2016}. These results are indicative that the repetitively cleaved \bise{} slab is composed of multiple slightly misaligned crystalline domains that accommodate a complex internal strain. 
This conclusion is further confirmed by the Lauegram obtained from the mechanically bent \bise{} single crystal [Fig.~\ref{fig:fig1}(g)], where the diffraction spots are elliptically distorted. 
The asymmetric shape of the streak in Fig.~\ref{fig:fig1}(f) reveals the variations in both the in-plane and out-of-plane lattice parameters, suggesting a competition between axial stretching and lateral contraction during strain relaxation and a presence of complex strain profile across the slab. 
Micro-Raman measurements [Fig.~\ref{fig:fig1}(h)] were employed to examine the distribution of the structural parameters across the \bise{} slab. The data reveal a complex lateral macrostrain profile ranging from $-0.9$\,\% to $+0.9\,$\% [Fig.~\ref{fig:fig1}(i), Supplementary Section~I.D]. These structural inhomogeneities likely originate from local elastic relaxation under non-uniform deformation upon cleavage, leading to domain-level variations of the lattice geometry within the macroscopic \bise{} crystalline slab.

\begin{figure*}
	\centering
	\includegraphics[width=0.9\linewidth]{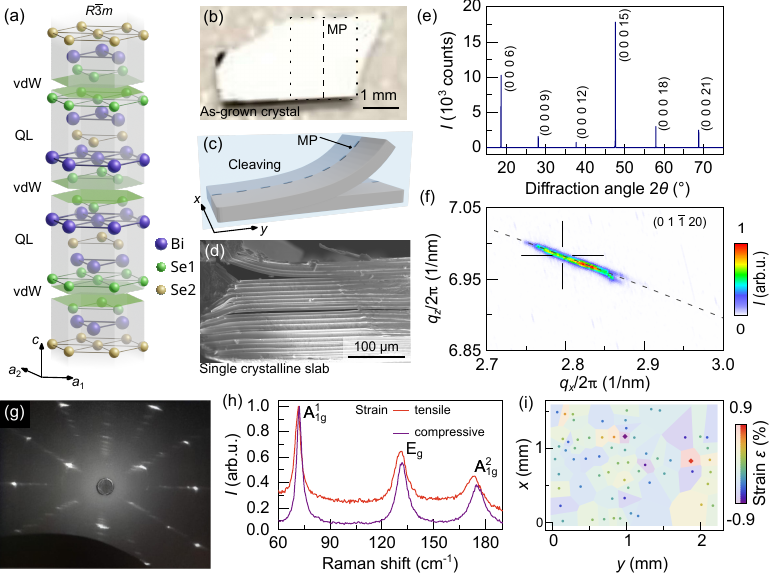}
	\caption{\textbf{Macroscopic \bise{} single crystalline slab.} 
	(a) Schematics of \bise{} crystal structure composed of quintuple layers (QL) that are interfaced by the vdW gaps.
	(b) Optical image of the pristine \bise{} single crystal used in this study. The black dashed line indicates the orientation of one of the mirror planes of the \bise{} crystal lattice and black dotted rectangle indicates the sample piece cut for transport measurements.
	(c) Schematics of the cleavage process. The mirror plane (indicated in blue) is oriented along the long side of the slab, which is repeatedly cleaved until the sample thickness is reduced to $200\,$µm.
	(d) SEM image of the side face of the cleaved \bise{} single crystalline slab. The image identifies layered structure of the crystal.
	(e) X-ray diffraction pattern of the \bise{} single crystal. 
	(f) Reciprocal space map of the asymmetric $(0\,1\,\overline{1}\,20)$ reflection. The black cross indicate the position of the reflection for strain-free \bise{} according to~\cite{Cavallin2016} and gray dashed lines show the projection profile for the corresponding reflections. 
	(g) Lauegram taken of the \bise{} single crystal after two cleavage iterations and under macroscopic bending.
	(h) Raman spectra measured of the cleaved \bise{} single crystal slab showing $A^1_{1g}$, $E_g$ and $A^2_{1g}$ modes. The spectra are taken at the location of the maximal compressive and tensile strains shown in panel (i) with diamond symbols.
    (i) Micro-Raman strain mapping revealing the strain distribution map across the \bise{} single crystal obtained from the frequency shift of the $A^2_{1g}$ Raman mode, the average uncertainty is estimated from peak fit error as $0.079$\,\%.
	}
	\label{fig:fig1}
\end{figure*}

\subsection*{Transport characterisation}

\begin{figure*}
	\centering
	\includegraphics[width=0.95\linewidth]{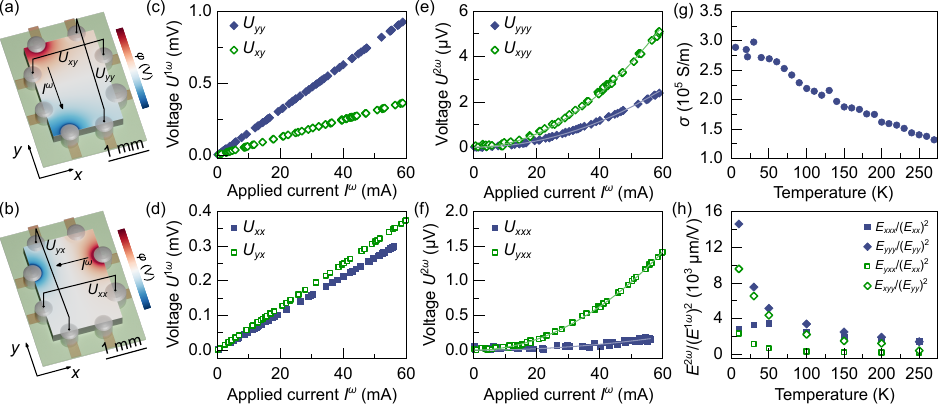}
	\caption{\textbf{Electrical transport characterisation of \bise{} single-crystalline slab.} 
	(a,b) Schematics of the \bise{} crystalline slab interfaced with 8 contacts for electrical transport measurements. 
	The surface of the crystal is color coded representing the distribution of the electrical potential across the crystal under electrical current sourced along the (a) long and (b) short edge of the slab [see also Supplementary Fig.~10].
	(c,d) The longitudinal and transversal voltages measured at $30$\,K at the (c,d) first harmonic and (e,f) second harmonic.
    Error bars denote standard error of the signal measured for $60$\,s. Lines in panels (e,f) are fits to the data accordingly to $U^{2\omega} \propto (I^\omega)^2$.
    (g) Temperature evolution of linear conductivity $\sigma$ of the \bise{}.
    (h) Temperature dependence of the longitudinal and transversal responsivities of the \bise{} slab.
	}
	\label{fig:fig2}
\end{figure*}

For the electrical transport measurements, the cleaved \bise{} single crystalline slabs were interfaced with 8 contacts along the perimeter [Fig.~\ref{fig:fig2}(a,b)] enabling the measurement of longitudinal and transverse responses in nonlocal geometry. Several samples subjected to different cleavage procedures were studied giving qualitatively similar results, as reported in Fig.~\ref{fig:fig2} and Supplementary Figs.~12 and 14. The representative current-voltage curves measured at 30\,K under different configurations of the driving first-harmonic current, $I^\omega$, are shown in Fig.~\ref{fig:fig2}(c,d). The corresponding temperature-dependent data are provided in Supplementary Fig.~12. Within the applied range of currents, we observe linear current-voltage characteristics on first harmonics. Bulk electrical transport measurements of the \bise{} crystal [Supplementary Fig.~9] reveal typical metallic behaviour with conductance dominated by electrons. We note that $U_{xy}$ is nonzero because of the nonlocal measurement geometry used to characterize the single crystalline slabs. This is evident from the electrical potential distribution across the slabs shown in Fig.~\ref{fig:fig2}(a,b), which results in a spatially non-uniform current flow and mutual signal leakage between the longitudinal and transverse channels. This effect is additionally illustrated by numerical simulations of the current density distribution across the slabs [Supplementary Fig.~10].
By matching numerical simulations of the linear transport to the experimental data, we confirm that the conductivity tensor of the \bise{} slab remains in-plane isotropic. The temperature evolution of the conductivity $\sigma$ of \bise{} is shown in Fig.~\ref{fig:fig2}(g). 

Representative current-voltage curves at second harmonic measured at $30$\,K are shown in Fig.~\ref{fig:fig2}(e,f). Pronounced transverse second harmonic responses, $U_{xyy}$ and $U_{yxx}$, are observed for both applied current directions, with the nonlinear signal exhibiting a clear quadratic dependence on the sourced current [lines in Fig~\ref{fig:fig2}(e,f)]. 

The nonlocal measurement geometry complicates the direct quantification of these nonlinear transport responses. Since the measurement electrodes are distributed along the perimeter of the sample [Fig.~\ref{fig:fig2}(a)], rather than in a straight line, as in a conventional 4-probe geometry, the driving electric field, defined as $E^{1\omega} = U^{1\omega}/L$, where $L$ is the distance between the corresponding contacts, is underestimated. This has impact on the quantitative assessment of the device responsivity, which is calculated as $E^{2\omega}/(E^{1\omega})^2$. To account for these effects of nonlocal transport measurement geometry, we extracted the local electrical current densities between the corresponding contact pairs [Supplementary Fig.~10] and utilized them as scaling factors to calculate the true nonlinear responsivities [Fig.~\ref{fig:fig2}(h)]. Note that because the longitudinal linear and nonlinear voltages are obtained from the same pair of electrical contacts, no additional scaling is required to determine the longitudinal responsivity.

The calculated temperature evolution of the nonlinear responsivity for the \bise{} slab is shown in Fig.~\ref{fig:fig2}(h). The maximum responsivity is obtained for the $U_{yyy}$ signal reaching remarkable $14\times10^{3}\,$µm/V at $10$\,K. As the temperature increases, the responsivity decreases, eventually vanishing at about $250$\,K.
Although a similar temperature evolution of the second-harmonic transport response in \bise{} thin films has been reported previously~\cite{He_2021}, our single-crystalline \bise{} slabs exhibit a substantial six-order-of-magnitude enhancement in responsivity compared to those epitaxial thin films [Table~\ref{tab:responsivity}]. We attribute this giant increase in responsivity not to quantum geometry, but to the classical geometric nonlinear transport effect enabled by the inhomogeneous strain within our samples [Fig.~\ref{fig:fig1}(e,f)].

The presence of the classical geometric effects can be inferred from the macrostrain distribution across the sample. Spatial variations in the macrostrain [Fig.~\ref{fig:fig1}(i)] imply that the mechanical neutral, i.e., strain-free plane becomes a curved surface~\cite{Ortix_2011}. The resulting curvature of the electron trajectories gives rise to the gNLTE, a phenomenon previously observed in transversal transport channel in graphene and Bi arc-shaped microstructures~\cite{Schade_2019,Makushko_2024}, but here induced within the bulk of a macroscopic crystal by an inhomogeneous strain distribution.

\begin{figure*}
	\centering
	\includegraphics[width=0.35\linewidth]{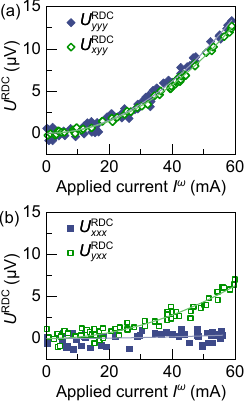}
	\caption{\textbf{Rectification performance of \bise{} slab.}
	Rectified DC voltage, $U^\text{RDC}$, measured at $30$\,K as a function of the AC current, $I^\omega$, applied along (a) long and (b) short sides of the \bise{} slab. 
	}
	\label{fig:fig3}
\end{figure*}

The technological potential of these macroscopic \bise{} slabs is further demonstrated by their efficient AC-to-DC conversion capability, which is a key requirement for zero-threshold energy harvesting. Fig.~\ref{fig:fig3} shows the rectified DC (RDC) signal recorded for the \bise{} slab at $30$\,K, with corresponding data at other temperatures provided in Supplementary Fig.~12. In agreement with the second-harmonic transport characteristics, the rectified DC signal scales quadratically with the applied AC current. 
A pronounced transverse rectified voltage is observed in both measured configurations, $U_{yxx}^\text{RDC}$ and $U_{yxx}^\text{RDC}$. Additionally, we detect a strong longitudinal $U_{yyy}^\text{RDC}$ signal and a slightly weaker $U_{xxx}^\text{RDC}$ response, consistent with the behavior of nonlinear responsivity of the \bise{} slab discussed above. 

\begin{figure*}
	\centering
	\includegraphics[width=0.35\linewidth]{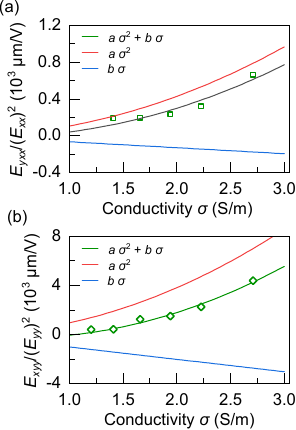}
	\caption{\textbf{Scaling analysis of nonlinear responsivity of single crystalline \bise{} slab.}
	The variation of transversal nonlinear responsivity of the \bise{} slabs with conductivity $\sigma$ can be fitted with $a \sigma^2 + b \sigma$ function.
	}
	\label{fig:fig4}
\end{figure*}

\subsection*{Scaling analysis}
We next 
develop
a scaling law for the gNLTE that is similar in essence to the scaling law for the disorder-driven nonlinear Hall effect~\cite{Du_2019,Makushko_2024}. We start by noticing that the gNLTE nonlinear resistance defined by the relation  $U_{ijj}=R_{ijj} I^2$ cannot directly depend upon the momentum relaxation time -- the effect is classical in nature and only related to classical electron trajectories. Additionally, a dimensional analysis shows that in terms of system volume, carrier density and effective mass of the carriers the most general form for the gNLTE nonlinear resistance
$$R_{ijj} \propto (n V)^{\alpha} \dfrac{m}{n e^3} \kappa.$$
Here, $\kappa$ is a characteristic length scale set by the strain gradient which ensures that $R_{ijj}$ cannot be associated with a uniform nonlinear conductivity. The specific value of $\alpha$ regulates instead the carrier density and volume dependence of the classical nonlinear resistance. It is interesting to note that the 
classical theory of the gNLHE in arc-shaped microstructures~\cite{Schade_2019,Makushko_2024} 
is recovered for $\alpha=-1$ when identifying the length scale $\kappa$ with the geometric curvature of the arcs.
To establish the scaling law for the gNLTE responsitivity, we further use that
the current $I \propto \sigma E_{jj}$ with $\sigma$ the linear conductivity.  
As a result,
we have that the gNLTE responsivity 
$$\dfrac{E_{ijj}}{E_{jj}^2} \propto n^{\alpha-1} V^{\alpha}  \dfrac{m} {e^3}{\kappa} \, \sigma^2.$$ The responsivity of the gNLTE is therefore expected to scale with the square of the linear conductivity assuming a fixed carrier density.

However, a subleading contribution that scales linearly with $\sigma$ can still exist. Its presence can be rationalized by noticing that the carrier density can display a weak temperature dependence of the form $n(T)=n_0 + \beta T$. From Hall resistance measurements at variable temperature [see Supplementary Fig.~9] we can evidence the occurrence of this feature in our \bise{} slabs. 
Using in addition that 
in the temperature range where the gNLTE is detected we can assume a temperature scaling of 
the linear conductivity $\propto T^{-1}$, we have that $n(\sigma)=n_0 + \gamma / \sigma$. As a result, the scaling law of the gNLTE responsivity  $E_{ijj}/(E_{jj}^2) = a \sigma^2 + b \sigma $. Crucially, the relative sign between the $a$ and $b$ coefficients is set by the temperature dependence of the carrier density (via the parameter $\beta$) and the temperature dependence of the nonlinear resistance (via the 
exponent $\alpha$). The fact that both the carrier density [see Supplementary Fig. 9] and the gNLTE responsitivity [see Fig. 2] decrease with temperature implies that $\beta<0$, while $\alpha>1$. Consequently,  
an opposite sign of the $a$ and $b$ coefficients 
is expected.
Our experimental data [Fig.~\ref{fig:fig4}] fulfill precisely these expectations for the scaling law.
This further verifies the dominant role of the classical gNLHE which in our \ch{Bi2Se3} slabs is activated by the inhomogeneous strain distribution.

\section*{Conclusion}

In summary, we presented an experimental study of nonlinear transport in single crystalline bulk topological insulator \ch{Bi2Se3}, which is inhomogeneously strained due to multiple mechanical cleavage steps. The samples reveal a record high  nonlinear responsivity exceeding $10^4\,$µm/V. Considering that the nonlinear responsivity of these bulk samples is six orders of magnitude larger than that observed in thin films of epitaxial \ch{Bi2Se3}, we attribute this major performance boost to the classical geometric nonlinear transport effect rather than transport phenomena
driven by the quantum geometry of the Bloch waves. 

The experimental observation of a giant rectified DC voltage in a bulk material holds implications for ambient electromagnetic energy harvesting. Unlike conventional semiconductor-based diodes, which suffer from a finite turn-on threshold voltage, the strain-induced geometric nonlinear 
transport
effect operates at zero bias and without any intrinsic threshold. This allows the macroscopic \ch{Bi2Se3} slabs to rectify weak high-frequency ambient signals that would otherwise remain unharvested. Furthermore, because the underlying mechanism relies on classical macroscopic strain gradients rather than fragile surface states, the rectification performance is robust and less susceptible to surface degradation. These characteristics establish macroscopic strain engineering of topological materials as a highly promising platform for scalable, high-efficiency broadband energy harvesters. 

We anticipate that  nonlinearities induced by inhomogeneous strain fields may be present in a broad class of layered materials that are highly susceptible to mechanical deformations. Ideal candidates are bulk materials with crystal structures featuring weak van der Waals bonds between layers, which facilitate the introduction of non-uniform strain upon mechanical manipulation, e.g., exfoliation or bending.
Three-dimensional topological insulators such as Bi$_2$Te$_3$ and Sb$_2$Te$_3$, which belong to the same family as the Bi$_2$Se$_3$ studied here, are natural candidates. Other layered semimetals and conductors with centrosymmetric bulk structures -- including certain polytypes of transition metal dichalcogenides (TMDs) like MoS$_2$ and WSe$_2$ -- are also promising, as their weak interlayer coupling enables significant and non-uniform local symmetry breaking. 
Furthermore, two-dimensional heterostructures and moir\'e systems, which are highly sensitive to hetero-strain and structural imperfections, provide an ideal playground to host these engineered nonreciprocal transport effects. 

\section*{Methods}

\subsection*{Crystal growth}
High purity starting elements (Bi, 99.997 \%, Alfa Aesar; Se, 99.999 \%, Alfa Aesar) weighed in stoichiometric ratio were placed in a quartz tube purged with argon for four times and then sealed in a vacuum of 10$^{-5}$\,mbar. After sealing, the quartz tube was positioned in a box furnace for three thermal treatments. For the first two steps, the temperature was increased from room temperature to 750°C at the rate of 250°C/h and kept for 12\,h. At the end of each thermal treatment, the samples were quenched to room temperature. For the third step, the temperature was increased from room temperature to 820°C at the rate of 250°C/h and kept for 24\,h. After that, the temperature was slowly decreased to 540°C with a rate of 10°C/h kept for 15\,h before quenching the sample again. To further improve the mixing and ensuring phase homogeneity, the sealed quartz tube was taken out of the furnace and turned upside down between each of the heat treatment steps. To achieve solidification of bigger crystals, quartz tubes containing synthesized \bise{} were placed vertically in a two-mirror optical image furnace (NEC Machinery, model SC1-MDH11020) where the material was melted again within a high thermal gradient using a very low translation speed (0.5\,mm/h). At the end of the growth, easily cleavable gray in colour large crystals with flat and shiny surfaces were obtained. Further details on the sample synthesis can be found elsewhere~\cite{Guarino2024}.

\subsection*{Structural analysis}
The structural properties of \bise{} were investigated using X-ray diffraction (XRD) techniques, including Bragg-Brentano geometry, parallel beam measurements, 2D pole figure analysis, and reciprocal space mapping (RSM). These methods provided detailed insights into the crystallographic orientation, lattice parameters, strain distribution, and overall quality of the sample. XRD studies were carried out using a Rigaku SmartLab 3\,kW with a parallel beam of Cu-K$_{\alpha1}$ radiation, $\lambda = 1.54056$\,\AA (using a Ge ($2\,2\,0$) two-bounce monochromator). For the measurements performed in the Bragg-Brentano geometry, the Cu-K$_{\alpha}$ mean radiation was used. The reflection positions and full width at half maximum (FWHM) of the Bragg-Brentano and parallel beam measurements obtained by the profile deconvolution was performed with WinCSD~\cite{Akselrud2014}.
XRD data were collected also by means of a Philips X'Pert-MRD high resolution diffractometer equipped with a four circle cradle. The incident beam is generated by X-ray tube focusing Cu-K$_{\alpha1}$ wavelength on the sample by an asymmetric four crystal Ge $(2\,2\,0)$ Bartels monochromator followed by a graded parabolic Guttman mirror. The diffracted beam reaches the detector with an angular divergence of $12$\,arcsec crossing a triple axis attachment and undergoing three $(0\,0\,2)$ reflections within a channel cut Ge crystal.
The Lauegrams were obtained in high resolution back-scattering Laue X-ray camera (Photonic Science,UK), the beam spot size on the sample was about $500$\,µm.

\subsection*{Raman spectroscopy}
The Raman spectra were measured using JY Horiba LabRAM HR Evolution Raman microscope in the backscattering geometry. The excitation wavelength was $532$\,nm with the incident power of less than $1$\,mW. The radiation is focused and collected by a $100\times$ microscope objective and the spot size on the sample’s surface is around $2$\,µm.

\subsection*{Sample preparation for magnetotransport measurements}
After synthesis, the single crystals were cleaved and cut in rectangular-shaped pieces with the long side parallel to the mirror plane, as shown in Fig.~\ref{fig:fig1}. The crystals were mounted on lithographically defined electrical contacts made of 35\,µm-thick \ch{Cu} on a mm-thick insulating FR4 substrate (Bungard Elektronik GmbH \& Co.KG, Germany). The contact layout was realized enabling the access to longitudinal and transversal resistances in a 4-point configuration with two orthogonal orientations of electrical current.
The crystals were mechanically fixed to the sample holder by GE-varnish glue (Oxford Instruments, UK) and contacted to the Cu pads using conductive silver paint (Plano GmbH, Germany). 
This contacting scheme allowed us to achieve long-term stability of the sample resistance, which is needed to perform reliable measurements of the linear and nonlinear transport. 

\subsection*{Bulk magnetotransport and sheet resistance measurements}
The pristine as-grown crystal is about 1-mm-thick, which makes it highly conductive. A typical resistance is in the mOhm range. In this regime, the measured signal is dominated by contact and wiring resistances, making a reliable nonlinear transport measurements impossible. To enable accurate transport measurements, we reduced the thickness by repeated cleaving, which increases the sample resistance to a measurable range of tens of mOhm.
Bulk transport measurements were carried out in a Lake Shore cryostat in the temperature range from 5 to 300\,K. The measurement was performed in the van der Pauw configuration using 4 electrical contacts positioned at the edges of the sample. For the Hall effect measurements, magnetic field of up to 5\,T was applied perpendicular to the sample surface.

\subsection*{Linear transport simulations}
Numerical simulations of the linear transport in \bise{} crystalline slabs were performed using Comsol Multiphysics 6.2 (Build: 415). The lateral size and the thickness of the model geometry was taken as in the experiment. The shape of electrical contacts was set as semicircles of the same size. The location of contacts was fixed at the edge of the sample. Slight lateral adjustments of the contact locations were performed to match the experimental and modelled voltages across the corresponding contact pairs. The conductivity of the contact pads in modeling was set to $1.2\cdot10^6$\,S/m. The electrical potential is applied to the top surface of the electrode, resulting in an even distribution of the potential on cold and hot electrodes, respectively. Any possible anisotropy of the electrical conductivity in the $z$ direction is neglected.

As our COMSOL simulations do capture the experimental data in linear transport with a homogeneous and isotropic in-plane conductivity tensor, we can exclude any carrier density and mobility gradients in our samples even after cleavage. We also note that the fact that the linear conductivity tensor is a scalar is compatible with the presence of a rotational symmetry for the global structure.

Joule heating effects are also estimated from the COMSOL model. The model assumes a \bise{} crystal slab to be in a direct contact with an FR4 substrate. The back side of the FR4 substrate acts as a heat sink with the heat transfer coefficient of $100$\,W/(m$^2$K). A heat transfer via radiation is neglected and convection into ambient (pressure $\approx 10^{-3}$\,mbar) is set to $0.1$\,W/(m$^2$K).
The maximum applied electrical current of $75$\,mA results in the heating of not more than $0.3$\,K with the maximal difference across the sample surface of $0.15$\,K (Supplementary Fig.~11). Therefore, we exclude the Joule heating as a potential source of the nonlinear transport signal.

\subsection*{Harmonic transport measurements}
The harmonic measurements reported in the main text were done using a Tensormeter measurement device (Tensor Instruments, HZDR Innovation GmbH, Germany). Single crystals of \bise{} are good conductors and reveal the 4-point resistance in the range of several tens mOhm only. The measurements were carried out in a 4-point configuration in a closed cycle cryostat (Easycool, CSIC PRIDe (Nanjing) Cryogenic Technology Co., LTD.) in the temperature range from 10 to 300\,K. 
The absence of nonlinearities of the measurement setup, assuring an accurate detection of both second harmonic and rectified DC signals, has been confirmed by reference measurements on a surface mounted device (SMD) resistor and SMD resistor-diode circuits [Supplementary Fig.~13].
In addition to the measurements with a Tensormeter device, we performed electrical transport harmonic measurements using Multifrequency Lock-in Amplifier MLA-3 (Intermodulation products, Sweden), Supplementary Fig.~15. 

In all transport measurements, the sinusoidal current at the fundamental frequency of $16.67$\,Hz was sourced using internal generator of the measurement device. In all plots, we report amplitudes of the measured longitudinal and transverse voltages (at DC, first and second harmonics). For each sourcing current value, the data was averaged over one minute integration time. In measurements with MLA-3 device the bandwidth was set to $0.33$\,Hz.

In addition to the sample discussed in the main text, we performed a similar set of measurements on another \bise{} sample with a different strain state [Supplementary Fig.~14]. The data is qualitatively similar to the one discussed in the main text of the manuscript. There are quantitative differences, which are related to the geometry of the contacts and different strain state of the \bise{} slab.

The nonlocal measurement geometry (electrical contacts are not positioned along a straight line, line in a Hallbar structure) leads to underestimation of the driving electric fields $E^{1\omega} = U^{1\omega}/L$, where $U$ is the measured voltage and $L$ is distance between the corresponding contacts. This impacts the assessed values of nonlinear responsivity calculated as
$E^{2\omega}/(E^{1\omega})^2 = U_{ijj}/(U_{jj})^2 L_j^2/L_i.$
To account for this effect we extract the local electrical current densities between the corresponding contacts [Supplementary Fig.~10] and implement their ratio as a correction factor $\delta = j_\text{DRV}/j_\text{SNS}$, with $j_\text{DRV}$ and $j_\text{SNS}$ being local current in between drive adn sensing contacts, accordingly. The nonlinear responsivities are then calculated as:
$$
\begin{aligned}
\frac{E_{iii}}{(E_{ii})^2} &= \frac{U_{iii}}{(U_{ii})^2} \frac{L_i^2}{L_i} \\
\frac{E_{ijj}}{(E_{jj})^2} &= \frac{U_{ijj}}{(\delta U_{jj})^2} \frac{L_i^2}{L_j}
\end{aligned}
$$
Here, $\delta = 0.268$ for current applied along $x$ direction and $\delta = 0.528$ for current along $y$. The nonlinear responsivity in longitudinal channel does not need the correction since both $U_{ii}$ and $U_{iii}$ are measure on the same pair of contacts.

\section*{Data availability}
All data that support the plots within this paper and other findings of this study are available from the corresponding authors upon reasonable request. Source data are provided with this paper.

\section*{Acknowledgements}
We thank Prof. Olav Hellwig (HZDR, TU Chemnitz) for providing access to the XRD tool. 

\section*{Funding}
This work is financed in part via the European Commission HORIZON projects (project REGO with ID: 101070066 and Marie Skłodowska-Curie Grant 101106524) and ERC grant 3DmultiFerro (project number: 101141331). 
C.O., R.B.A. and Deb.M. acknowledge support by the Italian Ministry of Foreign Affairs and International Cooperation PGR12351 (ULTRAQMAT) and from PNRR MUR Project No. PE0000023-NQSTI (TOPQIN). 

\section*{Author contributions}
A.V., C.O., and D.M. developed the project idea. The samples were prepared by A.V., R.T.F., A.G., and M.W.U.R.. Structural characterization was performed by A.V., R.T.F., A.G., M.W.U.R, F.G., and I.V.. I.V., P.M., and P.T.D. performed transport characterization with the support from T.K. and D.M.. Analysis of the transport data was done by P.M., I.V., D.M., C.O. with support from T.K., P.T.D., R.B.A.. R.X. and P.M. performed linear transport simulations with the support from D.M.. O.P. performed Raman characterization and data analysis.
O.A.P. performed Laue measurements with support from R.T.F., A.G., and A.V.
C.O., R.B.A., and Deb.M. developed theoretical description of the transport effects in strained \bise{} crystals. The manuscript was written by D.M., C.O., P.M., I.V., and A.V. with the contribution from P.T.D., T.K., R.T.F., A.G., M.W.U.R., R.X., F.G., R.B.A., O.P., and Deb.M.. 
	
\section*{Competing interests}
The authors declare no competing interests.

\end{document}